\documentclass[sigconf]{acmart}

\usepackage[show]{chato-notes}

 \renewcommand\footnotetextcopyrightpermission[1]{} 

\def\BibTeX{{\rm B\kern-.05em{\sc i\kern-.025em b}\kern-.08emT\kern-.1667em\lower.7ex\hbox{E}\kern-.125emX}}
    
\acmConference[FACTS-IR Workshop @ SIGIR]{FACTS-IR Workshop @ SIGIR '19: ACM SIGIR}{2019}{Paris, France}
\acmBooktitle{FACTS-IR Workshop @ SIGIR '19: ACM SIGIR, July 21--25, 2019, Paris, France}

\newcommand{\g}[1]{#1}

\begin{document}

\title[]{The FACTS of Technology-Assisted Sensitivity Review}

\author{Graham McDonald, Craig Macdonald, Iadh Ounis}
 \affiliation{%
   \institution{University of Glasgow}
   \city{Glasgow}
   \country{Scotland, UK}}
\email{firstname.lastname@glasgow.ac.uk}

\renewcommand{\shortauthors}{McDonald, et al.}

\begin{abstract}
At least ninety countries implement Freedom of Information laws that state that government documents must be made freely available, or \textit{opened}, to the public. However, many government documents contain sensitive information, such as personal or confidential information. Therefore, all government documents that are opened to the public must first be reviewed to identify, and protect, any sensitive information. Historically, sensitivity review has been a completely manual process. However, with the adoption of born-digital documents, such as e-mail, human-only sensitivity review is not practical and there is a need for new technologies to \textit{assist} human sensitivity reviewers. In this paper, we discuss how issues of \textit{fairness}, \textit{accountability}, \textit{confidentiality}, \textit{transparency} and \textit{safety} (FACTS) impact technology-assisted sensitivity review. Moreover, we outline some important areas of future FACTS research that will need to be addressed within technology-assisted sensitivity review.
\end{abstract}

\maketitle

\section{Introduction}\label{sec:intro}
At least ninety countries around the world implement Freedom of Information (FOI) acts, i.e. laws that state that information that is produced by the governments or public bodies must be made available, or \textit{opened}, to the public~\citep{peled2010constitutional}, for example in the UK~\cite{foia} and the USA~\cite{usfoi}. FOI assumes that all of the information within government documents will be made available, either through a \textit{release on request} model whereby documents are opened in response to individual FOI requests~\cite{foia,usfoi} or an \textit{open by default} model whereby historical documents are transferred to a public archive~\cite{pra1958}. However, government documents can contain \textit{sensitive} information, such as information that would likely damage the national security or international relations of countries if the information was made freely available. Therefore, FOI acts typically provide \textit{exemptions} that negate the obligation to open information that is of a sensitive nature. Moreover, all government documents that are opened to the public must first be manually \textit{sensitivity reviewed} to identify and protect any sensitive information.  

Over the last twenty to thirty years, governments have increasingly used born-digital documents, such as emails, word processing documents, PDFs and on-line discussions, instead of paper documents. These digital documents now need to be made available to the public and, therefore, must be sensitivity reviewed. It has been widely recognised, e.g. in ~\cite{darpa,digitalrecordsreview,naTarBornDigital,digitalstrategy}, that \g{due to the volume of documents that are to be reviewed} there is a need for technology-assisted review (TAR) approaches: i.e. using information retrieval (IR) and machine learning technologies to identify sensitive information and {\em assist} with the sensitivity review of digital documents, this is usually referred to as \textit{Technology-Assisted Sensitivity Review}. 

Recently, advances have been made in developing automatic approaches for classifying sensitive information and assisting digital sensitivity review, e.g.~\cite{mcdonald_ECIR2014,mcdonald_ECIR2018als}. However, to adhere with FOI and due to the potential consequences of releasing sensitive information, assistive technologies will need to strictly adhere to principles of \textit{fairness} (e.g. to not discriminate against particular groups of people), \textit{accountability} (e.g. produce results in a principled manner), confidentiality (e.g. not reveal sensitive or secret information), \textit{transparency} (e.g. by explaining their decisions) and \textit{safety} (e.g. mitigate against adversarial abuse) (FACTS)~\cite{culpepper2018research}.   

In this position paper, we present our current views on how issues relating to FACTS have specific implications for IR and TAR systems that are explicitly tasked with finding, or assisting a human reviewer to find, sensitive information. Moreover, we discuss some of the more pressing areas of future research that will enable IR and TAR systems are to be able to maintain FACTS in technology-assisted sensitivity review.

The examples that we discuss in this paper focus on the UK government domain. However, there is a need for technology-assisted sensitivity review in many other countries (and for other organisations, such as police departments). For example, the expectation within the archival community is that Commonwealth countries are likely to follow what the UK implements. Indeed, a recent study has found that there is already a real demand for technology-assisted sensitivity review in Malawi~\cite{tough2018scope}.

The remainder of this paper is as follows: We begin by providing a short overview of technology-assisted sensitivity review and related tasks in Section~\ref{sec:tasr}. In Section~\ref{sec:facts}, we discuss the implications of fairness (Section~\ref{sec:fair}), accountability (Section~\ref{sec:Accountable}), confidentiality (Section~\ref{sec:Confidential}), transparency (Section~\ref{sec:Transparent}) and safety (Section~\ref{sec:Safe}) for technology-assisted sensitivity review. Moreover, we provide an overview of what we argue to be some of the most important challenges that need to be addressed when developing responsible IR and TAR systems for automatically identifying sensitive information. Finally, we summarise our conclusions in Section~\ref{sec:conclusions}.

\section{Technology-Assisted Sensitivity Review and Related Tasks}\label{sec:tasr}
\begin{figure*}[t]
	\centering
	\includegraphics[width=0.9\textwidth]{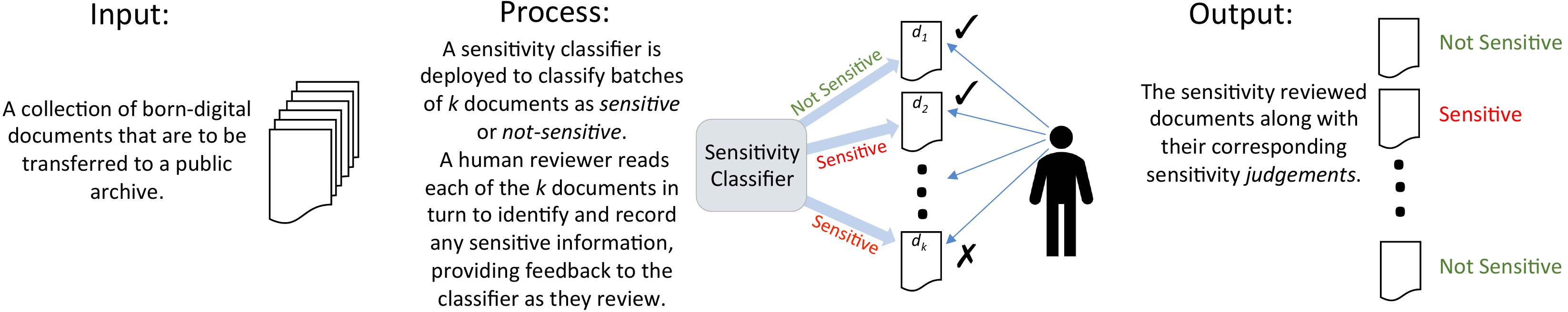}
	\caption{Technology-assisted sensitivity review: input, process and output.}
	\label{fig:dsr}
\end{figure*}

Figure~\ref{fig:dsr} illustrates the input, process and the output of the technology-assisted sensitivity review task. The input to the process is a collection of digital documents, $D$, that are to be opened to the public, e.g. that are to be publicly archived. Technology-assisted sensitivity review is an iterative process, whereby a sensitivity classifier is deployed to classify documents, $d_i \in D$, in batches of \textit{k} documents. A human reviewer then reads the \textit{k} documents and records a \textit{sensitivity judgement}, $j_i$, for each document. A sensitivity judgement, $j_i$, needs to record if the document is \textit{sensitive} or \textit{not sensitive}. For a document that is judged as sensitive, the sensitivity judgement must also include: (1) the sensitivities that apply to the document; (2) which parts (e.g. sentences or paragraphs) of the document they apply to; and (3) a textual description, or justification, about why the information is sensitive and should, therefore, be closed. The output of the digital sensitivity review process is the collection of sensitivity reviewed documents and the corresponding set of sensitivity judgements, $J$, where for each document, $d_i$, there is a sensitivity judgement, $j_i$. The information recorded in a sensitivity judgements, along with the newly judged documents, provide feedback to the sensitivity classifier after each iteration.

As can be seen from Figure~\ref{fig:dsr}, one of the main tasks for a technology-assisted sensitivity review system is to automatically classify documents by whether they do or do not contain sensitive information. Historically, most of the work on automatically classifying sensitive information in documents has focused on the anonymisation of structured \textit{personal data}, such as names, addresses, telephone or bank account numbers. Early examples of automatic anonymisation or redaction of documents used dictionaries (or medical knowledge bases) to term-match known sensitive terms and regular expressions to identify repeated patterns such as postal codes and dates of birth \citep{gupta2004evaluation,neamatullah2008automated,sweeney1996replacing}. Named entity recognition (NER) has also become a popular approach for \textit{masking} personal data, relating to \textit{persons}, \textit{organisations} and \textit{locations} ~\cite{DBLP:conf/mdai/AbrilNT11,cumby2011machine,dernoncourt2017identification,gardner2008hide,guo2006identifying,uzuner2008identifier}.\

It is only relatively recently, however, that research has advanced in the more general task of automatically classifying sensitive information that is exempt from public release through FOI. For example, in~\cite{mcdonald_ECIR2014}, we showed that text classification~\cite{Sebastiani:TC} is a viable approach for developing automatic sensitivity classification. Our initial research focused on learning sensitivity classification models from examples of historical sensitivities (i.e. \textit{offline} learning). More recently, however, our work has focused on an assistive model to aid the accuracy and speed of the sensitivity reviewer, whereby the classifier learns from a reviewer as they review (i.e. \textit{active} learning) to make predictions about sensitivities that are specific to the collection being reviewed~\cite{mcdonald_ECIR2018als}. Each of these tasks, i.e. learning from historical sensitivities, learning from reviewers and predicting sensitivity, have FACTS implications that will have to be addressed in future research. We discuss some of these implications in Section~\ref{sec:facts}.

Out-with the government context, the most closely related task to sensitivity identification of which we are aware is the identification of \textit{privileged} information in eDiscovery~\cite{oard2010evaluation}. In general, eDiscovery is concerned with the identification of documents that are responsive (relevant) to a production request in a lawsuit. Typically, less-senior paralegals are tasked with reviewing documents to identify responsive ones. Increasingly, active learning classification systems~\cite{settles2012active} are used to refine the rankings of documents to increase recall while reducing reviewer efforts~\cite{cormack2014evaluation}. However, before any responsive documents can be provided to the other party, each document must be reviewed to identify {\em privileged} information that should not be handed over: e.g. information that is protected due to solicitor-client privilege or potential self-incrimination. Typically, such privilege decisions are nuanced and a senior lawyer involved in the process will make them~\cite{linzy2011attorney}. Overall, privilege can be seen as a form of sensitivity and we expect that identifying privileged information will share many of the same FACTS implications as digital sensitivity review.

\section{The FACTS of Technology-Assisted Sensitivity Review}\label{sec:facts}
In this section we discuss some of the issues relating to fairness (Section~\ref{sec:fair}), accountability (Section~\ref{sec:Accountable}), confidentiality (Section~\ref{sec:Confidential}), transparency (Section~\ref{sec:Transparent}) and safety (Section~\ref{sec:Safe}) that future research on technology-assisted sensitivity review will need to address.

\subsection{Fairness}\label{sec:fair}

It is generally accepted that automatic approaches for identifying sensitive information (e.g.\ sensitivity classification~\cite{mcdonald_ECIR2014}), and technology-assisted sensitivity review~\cite{mcdonald_ECIR2018als}, will not completely replace human sensitivity review for the foreseeable future~\cite{naTarBornDigital}. However, as governments come to rely more on automatic technologies to help to decide what documents should or should not be released to the public, there is a risk that biases are introduced to the decision making process due to the classification learning process. For example, a sensitivity classifier that inaccurately learns that \g{members of a particular ethnic or societal group} are more likely to be terrorists \g{could potentially lead to a disproportionate number of closure applications\footnote{In the UK, if a document is judged as being sensitive, the government department must submit an application to a governing body to seek for the information to be redacted or the document to be retained. There are publicly available statistics on the closure applications that are made.} for documents relating to those groups.} Not only would the introduction of biases such as these have implications for the (groups of) people who are affected \g{(e.g. an incorrectly reinforced perception of an association with terrorism)}, but they also have the potential to result in decision making practices that are not legally sound. For example, in the UK, the Equality Act 2010~\cite{eqact} legally protects the rights of people from discrimination due to, for example, their age, race, disability, gender, religion or belief, gender reassignment, sex or sexual orientation.  

To mitigate against fairness issues such as these, it will be necessary to ensure that, when learning to identify sensitivity from historical examples, the real-world distributions of categories of groups (i.e. demographics) that are potentially likely to be discriminated against will have to be accurately reflected in the data that is used to train and evaluate the sensitivity classifiers. \g{Moreover, efforts will have to be made to mitigate against the representations, e.g. classification features, being constructed in a manner that introduces bias.} As a safety check, the sensitivity predictions that are made in production environments will need to be monitored and analysed to identify any skewed distributions in the predictions.

\subsection{Accountability}\label{sec:Accountable}
Government departments, or other public authorities that are responsible for releasing information through FOI, are held legally accountable for the decisions that they make when choosing to release, or not to release, documents. Therefore, for governments to be able to make use of assistive technologies for identifying sensitive information, \g{the technologies must be predicated on the same frameworks and guidelines that the government works within.} For example, the sensitivity predictions made by a sensitivity classifier must be guided by the details of the FOI laws. Moreover, sensitivity classifiers should not produce unreasonable or unexpected results. 

To be able to meet the accountability requirement, technologies for assisting digital sensitivity review must be developed in a principled manner to ensure that the steps that led to a specific sensitivity classification prediction being made can be identified. To address this, our approach so far has been to focus our research on developing \textit{linear} classification models in which each of the feature engineering approaches that we deploy allow for the decision making process to be traced, from a document's vectorised representation, back through the system to know precisely why the prediction was made. 

In recent years, however, there have been many advances in the state-of-the-art for text classification techniques, beyond linear classification models. It is expected that such approaches, e.g. CNNs~\cite{Zhang2015CharacterlevelCN} or Hierarchical Attention Networks~\cite{yang2016hierarchical}, may become more effective for sensitivity classification as the volume of documents that need to be sensitivity reviewed increases in the coming years. However, it has previously been shown that deep learning approaches, such as neural networks, can encode sensitive information into the learned model early in the learning process and this information can potentially be reverse engineered~\cite{secretSharer}. This is problematic since governments are likely to be legally obliged to release publicly certain details of the sensitivity review process. Therefore, for governments to adopt deep learning sensitivity classification approaches, there will need to be further research into the potential risks of inadvertently releasing sensitive information due to maintaining an expected level of public accountability.

Another consideration relating to the accountability of technology-assisted sensitivity review arises from the fact that sensitivity evolves over time. This could be due to a number of reasons. For example, it may be due to: changing political circumstances (e.g. the need for Brexit negotiations means that some previously made comments could now be more sensitive than if the UK was not leaving the EU); changes in the FOI laws that governments have to adhere; or from an individual exercising their right to be forgotten~\cite{kalis2014google}. To address these concerns, there is a need for new research into how sensitivity classification models that rely on historical data can retain expected levels of accuracy while remaining aligned with changing perceptions of sensitivity over time.

\subsection{Confidentiality}\label{sec:Confidential}
The primary purpose of technology-assisted sensitivity review is to \textit{assist} human reviewers to review and release to the public as many non-sensitive documents as possible in the least amount of time. As part of this process, sensitivity classification is tasked with identifying sensitive information. Therefore, the overall purpose of technology-assisted sensitivity review is to not release sensitive (or confidential) information when releasing documents to the public. Concerns about accidentally releasing such sensitive information are clear and we discuss them throughout the rest of this paper. 

There are, however, potentially other confidentiality matters that should be considered when developing technology-assisted sensitivity review systems. For example, when deploying active leaning approaches to learn the from reviewers, the reviewers actions must be logged as they perform their review. There is a potential for such logging activities to release information about the reviewers if the log data is not handled appropriately. Moreover, there is a potential risk that learning from reviewers could encode unconscious biases into a sensitivity classifier's model. If these biases skew the classifier's predictions to unfairly bias a particular group of people, it is possible that (due to the necessity for accountability of the system) certain biases could potentially be traced back to individual reviewers \g{if an \textit{in-house} audit was conducted}. If this were to happen, a reviewer may find themselves in an unfavourable situation. Therefore, technology-assisted sensitivity review systems will need to provide some reassurances that the data the system collects from the reviewers could not be used to their detriment.

\subsection{Transparency}\label{sec:Transparent}
Transparency is one of the founding principles of FOI. Ultimately, democratic governments must be seen to be transparent in their decision making process and TAR systems for sensitivity review are in effect decision-support systems. Moreover, in our previous work~\cite{McDonald_CHIIR2019}, we showed that providing reviewers with automatic sensitivity classification predictions can enable the reviewers to make accurate reviewing decisions more quickly.  

In practice, when assisting digital sensitivity review, this means that reviewers and governments departments must be provided a mechanism to reason about, and explain, \textit{why} the classifier makes each of its sensitivity predictions. As we previously stated in Section~\ref{sec:Accountable}, our research so far has focused on developing linear classification models. This approach, along with a SVM classifier for text classification, has enabled us to use a simple heuristic, i.e. a term feature's perpendicular distance from the classification hyperplane~\cite{guyon2002gene}, to provide reviewers with information about the terms in a document that are most influential in a sensitivity classification prediction. However, if more complex models are going to be deployed for sensitivity classification then a more principled approach will be necessary. There have been some recent examples of how the decisions of non-linear models can be explained, for example in~\cite{ribeiro2016should}. However, there will need to be further research into how approaches for explaining sensitivity predictions actually benefit sensitivity reviewers, and what types of explanations are most useful in practice.  

\subsection{Safety}\label{sec:Safe}
Current research into developing algorithms for classifying sensitive information, and assisting sensitivity review, has the intent of enabling the release of more non-sensitive documents into the public domain than would be possible through a purely manual approach. However, the sensitivity review process is not infallible and it will continue to be the case that some sensitive documents will be released accidentally. Therefore, developing sensitivity classification technologies does raise some concerns about how the technologies could be used for adversarial purposes.

Publicly archived digital government documents are more easily discoverable than their paper equivalents due to the ubiquity of modern indexing and search technologies (discovering archived paper documents requires someone to physically visit an archive and manually inspect a document collection.) There is a risk that sensitivity classification technologies could be used for adversarial purposes such as blackmail if the algorithms were to be acquired by people who are intent on using them for personal gain. This could clearly be damaging for whoever is being targeted and could potentially lead to governments overturning FOI laws. This, in turn, \g{would have implications for the ability of the public to scrutinise the actions of governments}. Future research will, therefore, have to evaluate the potential risk from such practices and try to ensure that technologies that are intended to protect sensitive information are not used for adversarial purposes.

\section{Conclusions}\label{sec:conclusions}
The need for technologies to assist with the sensitivity review of digital government documents has been recognised in many countries, including the UK and the USA. However, such technologies will have to ensure that they abide by principles of \textit{fairness}, \textit{accountability}, \textit{confidentiality}, \textit{transparency} and \textit{safety} (FACTS).  In this paper, we have discussed the implications for technology-assisted sensitivity review with respect to these principles and outlined some of the areas of future research that will need to be addressed. For example, developing methods to ensure that sensitivity classification does not unfairly discriminate against certain groups of people, and methods to minimise the risk of inadvertently releasing sensitive information while maintaining an expected level of public accountability. The demand for technology-assisted sensitivity review will continue to increase in the coming years and the principles of FACTS will be integral to the technology's success.

\bibliographystyle{ACM-Reference-Format}
\bibliography{bibliography}
\end{document}